\documentclass[11pt,a4paper]{article} 
\usepackage{jheppub}
\usepackage[utf8]{inputenc} 
\usepackage{graphicx} 
\usepackage{booktabs} 
\usepackage{array} 
\usepackage{paralist} 
\usepackage{verbatim} 
\usepackage{bbold}
\usepackage{slashed}
\usepackage[draft]{fixme}
\usepackage{subfigure}
\usepackage{float}
\graphicspath{{./}}
\usepackage{fancyhdr} 
\usepackage[nottoc,notlof,notlot]{tocbibind} 
\usepackage[titles,subfigure]{tocloft} 

\subheader{}
\title{Limit setting procedures and theoretical uncertainties in Higgs boson searches}
\author{Falko Dulat,}
\author{Bernhard Mistlberger} 
\affiliation{Insitute of Theoretical Physics, ETH Zürich, Switzerland}
\emailAdd{dulatf@itp.phys.ethz.ch}
\emailAdd{bmistlberger@phys.ethz.ch}
\abstract{We review the $CL_S$ method used by ATLAS and CMS to set exclusion limits on the mass of the Higgs boson. We investigate the impact of theoretical uncertainties of Higgs boson cross-sections due to higher order perturbative corrections and convolution with parton densities on exclusion limits. We illustrate the impact on limits of commonly used treatments of these uncertainties with a simulated Higgs search.  We adopt a "worst-case scenario" for the perturbative uncertainties. We find that the statistical methods of ATLAS and CMS do not incorporate this specific uncertainty adequately.}
\notoc
\begin{document}
\maketitle
\newpage
\section{Introduction}
The Large Hadron Collider has made significant progress searching for the Higgs boson in 2011. With the latest results from ATLAS \cite{Collaboration2012} and CMS \cite{Collaboration2012a} together with the electroweak precision data from LEP \cite{ALEPHCollaboration2010} only a very small mass window in the region of 125 GeV is not excluded for the standard model Higgs.

The statistical procedure that is used to exclude the existence of a Higgs boson is the $CL_S$ method \cite{ReadCLS}. The method can account for systematic and theoretical uncertainties. A specific concern that was voiced in many discussions \cite{Kilminster2012,Dawson2012} in the theoretical community is the correct treatment of perturbative and parton density uncertainties in this method. In this article we investigate whether the limits obtained with the $CL_S$ method can be considered sufficiently conservative regarding these uncertainties.

The paper is organized as follows. In section \ref{sec:cls} we present a general review of the $CL_S$ method, which is used to calculate exclusion limits in standard model Higgs boson searches. Next, we introduce the concept and use of nuisance parameters. In section \ref{sec:nuisance} we analyze the worst-case and profiling method used to treat nuisance parameters in limit setting procedures. In section \ref{sec:combo} we illustrate the results obtained in the previous sections and support our conclusions with a  exemplary simplified Higgs boson search. Furthermore, we use our example to discuss the common methods of presenting results of Higgs searches. Finally we make a recommendation for the treatment of perturbative uncertainties.
\section{Review of the $\mathbf{CL_{S}}$ method}
\label{sec:cls}

In high energy particle physics a theoretical model or hypothesis is usually represented by a quantum field theoretical Lagrangian density. Experiments such as the LHC test whether these theoretical models are suited to describe observations. Fundamental parameters of the Lagrangian, like coupling strengths or masses, can be extracted by fitting the quantitative predictions of  the model to experimental data.
Experimental results may have discerning power for different models that aspire to explain the same phenomena.
The $CL_S$ method \cite{ReadCLS},  as used by ATLAS and CMS \cite{atlashiggscombo,CMS-PAS-HIG-11-032},  provides a means of setting upper limits on cross-sections derived from a model and constrains the possible range of fundamental parameters. 

Cross-section predictions for a specific experiment depend on input parameters, i.e. fundamental parameters of the Lagrangian as well as derived parameters such as scales, parton density functions and detector efficiencies. These parameters can be classified as either nuisance parameters or parameters of interest. A nuisance parameter is any parameter that is not under investigation in an experiment but still has an impact on the predictions. Conversely, parameters of interest are the parameters that are being constrained in the respective analysis.
The Higgs sector of the standard model does not contain a free parameter that affects the cross-sections of processes involving a Higgs boson of a certain mass, leading to unique predictions at a fixed order in perturbation theory. Consequently, there is no fundamental parameter that could be constrained to set a limit on the existence of a Higgs boson.
For testing the model, the experimental collaborations have introduced a deformation of the prediction, scaling the Higgs cross-section with an arbitrary factor $\mu$ that is used as parameter of interest \cite{atlashiggscombo,CMS-PAS-HIG-11-032}. This parameter is referred to as signal-strength modifier. The $CL_S$ method is then used to constrain the value of $\mu$ \cite{ReadCLS}. The standard model Higgs boson hypothesis for a certain Higgs mass is said to be excluded, if $\mu$ is found to be less than one with a certain confidence. 
We point out that this procedure entails some conceptual problems.
\begin{itemize}
\item A separation between Higgs (signal) and non-Higgs (background) contributions is only valid if the Higgs width is small with respect to its mass.  For low Higgs masses, $m_h~<~400~\text{GeV}$, such an approximation is valid, as interference terms can safely be neglected. However, in the high Higgs mass regime, $m_h > 400 \text{GeV}$, sizable corrections \cite{Anastasiou2012} have to be taken into account. In general, these corrections will have a different scaling behavior with $\mu$ than the signal contributions.
\item The standard model does not provide a free parameter that could be used to uniformly rescale the cross-section of processes involving the Higgs boson. One way to introduce a physical parameter of interest is to consider more complex Higgs models, such as the strongly interacting light Higgs model \cite{Giudice2007}, which contains additional free parameters. The cross-section should be evaluated consistently from the model, taking into account non-linear scaling of the cross-section from higher order corrections.
\end{itemize}
Model predictions for hadron colliders are hindered by several obstacles.
\begin{itemize}
\item Observables are obtained using a perturbative expansion. This introduces an uncertainty due to missing higher order corrections. 
\item Parton distribution functions cannot be extracted from perturbation theory and are obtained from complementary measurements. Consequently, the statistical and systematic uncertainties of these measurements must be propagated to any calculated cross-section.
\end{itemize}
The results of theoretical calculations are commonly stated in combination with the corresponding perturbative uncertainty and an overall parton density function uncertainty \cite{Watt2012,TheNNPDFCollaboration2011}.
Theoretical cross-section uncertainties must be propagated through limit-setting procedures. In the $CL_S$ method they are taken into account using nuisance parameters. In the section \ref{sec:llr} the treatment of nuisance parameters in the CLs method in general is discussed. In section \ref{sec:nuisance} we discuss the use of nuisance parameters for the specific uncertainties stated above.

Experimental collaborations use additional nuisance parameters to account for experiment specific uncertainties, such as detector efficiencies. The focus of this article is on theoretical uncertainties and we will neglect other uncertainties and their nuisance parameters.

\subsection{Likelihood ratios}
\label{sec:llr}
Let $X = \{X_i\}$ be a set of measurements and $g(X_i|\mu, \nu)$ be the probability density for the observation $X_i$ according to a model with parameters of interest $\mu$ and nuisance parameters $\nu = \left\{ \nu_j \right\}$. Then the probability to measure the values $X$ in a single experiment is
\begin{equation}
P = \prod_{i=1}^{N} P(X_i) = \prod_{i=1}^{N} g(X_i|\mu,\nu) dX_i =L(X|\mu,\nu)\prod_{i=1}^{N} dX_i,
\end{equation}
where $P(X_i)$ is the probability to find a measurement in an infinitesimal interval around $X_i$.
The quantity
\begin{equation}
L(X|\mu,\nu) = \prod_{i=1}^{N} g(X_i|\mu,\nu),
\label{eq:L}
\end{equation}
which appears as a joint probability density function, is the likelihood \cite{Fisher,Reid2003,Berger1988,Berger1999,Barndorf-Nielsen1988,Kalbfleisch1985,Reid2010}.

In the case of a collider experiment, $X_i$ is the number of observed events in an experimentally defined region $i$ of the phase-space. The cross-section for an event that the experiments are interested in is very low compared to the total cross-section of any scattering event happening. Therefore the number of interesting events is Poisson distributed and $g(X_{i}\mid \mu,\nu)=e^{-\lambda_i} \lambda_i^{X_i}/X_i!$ is the Poisson probability of observing $X_{i}$ events given the model prediction  $\lambda_i=\lambda_i(\mu,\nu)$. The functional dependence of $g$ on $\lambda_i$ is not written explicitly as we want to stress the implicit dependence on $\mu$ and $\nu$. The set of predictions for all phase-space regions is in the following denoted by $\lambda(\mu,\nu)=\{\lambda_i(\mu,\nu)\}$.

In order to compute $\lambda(\mu,\nu)$ all input parameters are set to specific values. The value of a nuisance parameter is usually constrained by independent experiments or by physical assumptions.
These constraints are modeled as outcomes of an auxiliary measurement by defining the likelihood $\pi(\tilde{\nu}|\nu)$, with a predicted value $\nu$, of the outcome that assigns a probability to every value $\tilde{\nu}$ of a nuisance parameter.
If a nuisance parameter, such as the parton density function uncertainty, has been obtained from a real experiment the confidence intervals and associated distributions are used to determine $\pi$.
For an unmeasured parameter such as the perturbative uncertainty, the distribution $\pi$ has to be modeled from assumptions.

In the $CL_S$ method we define the likelihood for a specific outcome of a measurement as the product of the likelihood to observe the measured number of events $X$ times the likelihood to observe a nuisance parameter value $\tilde\nu$ in an independent experiment.
\begin{equation}
\mathcal{L}(X|\mu,\nu)=\prod_{i}\frac{ e^{-\lambda_i(\mu,\nu)} \lambda_i^{X_i}(\mu,\nu)}{X_i! }\times \prod_{j}\pi_j(\tilde{\nu}_j|\nu_j) \label{eq:priorL}.
\end{equation}
The left product runs over all regions of the phase-space, while the right product enumerates all nuisance parameters used.
We drop the dependence on $\tilde\nu$ on the left-hand side of \ref{eq:priorL} and for this section fix $\tilde\nu$ to the default value of the parameter, e.g. the world-average for a parameter that has been measured in independent experiments or the most plausible value for a theoretical parameter. We explain below how to obtain $\nu$ in the $CL_S$ method. 

We assess the agreement of the measurements $X$ with one prediction $\lambda(\mu,\nu)$ relative to the agreement with another prediction $\lambda(\mu',\nu')$ and therefore define the likelihood ratio (LR).
\begin{equation}
LR = \frac{\mathcal{L}(X|\mu,\nu)}{\mathcal{L}(X|\mu',\nu')}.
\end{equation}

A quantity, such as the likelihood ratio, that distinguishes two predictions based on their agreement with a set of data is referred to as test-statistic.
In general there are multiple ways of defining the test-statistic used in the $CL_S$ method \cite{Lyons2008,Gross:1099994} that differ from analysis to analysis. In this work, we follow the prescription by ATLAS and CMS outlined in \cite{atlashiggscombo,CMS-PAS-HIG-11-032} and define our test-statistic, called profiled log-likelihood ratio (LLR), as
\begin{equation}
\label{eq:llr}
q_{\mu}(X)= \left\{ \begin{array}{ll}
	 -2\log\frac{\mathcal{L}(X|\mu,\hat{\nu}_{\mu})}{\mathcal{L}(X|\mu',\hat{\nu}_{\mu'})} \ &, \ \mu \geq \mu' \geq 0 \\
	 0 \ &, \ \text{else}
	 \end{array}\right.
\end{equation}
We choose the nuisance parameters in this definition such that the likelihood of the observation is maximized. This means we obtain $\hat{\nu}_{\mu}$ by performing a constrained maximum likelihood fit, i.e. by finding $\hat{\nu}_{\mu}$  for a given $\mu$ such that 
\begin{equation}
\mathcal{L}(X|\mu,\hat{\nu}_{\mu}) \geq \mathcal{L}(X|\mu,\nu) \ \ \ \forall \ \ \nu.
\label{eq:fit1}
\end{equation}
$\mu^{\prime}$ is the value of the parameter of interest that corresponds to the global maximum of the likelihood
function given $X$, i.e. $\mu'$ and $\hat{\nu}_{\mu'}$ are obtained such that
\begin{equation}
\mathcal{L}(X|\mu',\hat{\nu}_{\mu'}) \geq \mathcal{L}(X|\mu,\nu) \ \ \ \forall \ \ \mu,\nu.
\label{eq:fit2}
\end{equation}
The profiled LLR is a positive quantity. The larger it is the more $X$ disagrees with the prediction $\lambda(\mu,\hat\nu_\mu)$ compared to $\lambda(\mu',\hat\nu_{\mu'})$.
To ensure that only positively scaled cross-sections are considered, we constrain $\mu'$ to be non-negative. The fact that the test-statistic is defined to be non-zero only for $\mu'\leq\mu$ implies that the limits on the signal strength are one-sided, i.e. only upper limits are considered.

\subsection{Probability density function of the test-statistic}
\label{sec:fpdf}
As the outcome $X$ of a measurement is subject to statistical fluctuations $q_{\mu}(X)$ will assume different values in independent measurements. The distribution of these values, assuming that the prediction $\lambda(\mu,\hat{\nu}_{\mu})$ describes the expectation value of the measurement outcome $X$,  is referred to as $f(q_{\mu}|\mu,\hat{\nu}_{\mu})$. 

Analytic evaluation of $f(q_{\mu}|\mu,\hat{\nu}_{\mu})$ is in general impossible. One way to approximate $f$ is to evaluate $q$ for a large number of simulated toy measurements or replicas \cite{atlashiggscombo}. 
First, a replica measurement $X^R$ is obtained by Monte-Carlo generating Poissonian random numbers $X^R_i$ for every considered phase-space region $i$ with an expected number of events $\lambda_i(\mu,\hat\nu_\mu)$. We find the values of the nuisance parameters used to derive $\lambda_i(\mu,\hat\nu_\mu)$ by performing the fit of equation \ref{eq:fit1} to $X$ for a given value of $\mu$.
Next we consider the statistical fluctuation in the auxiliary measurement of the nuisance parameters as in \cite{CMS-PAS-HIG-11-032}. Therefore we generate random numbers $\tilde\nu^R$ distributed according to $\pi(\tilde{\nu}^R|\hat\nu_\mu)$  using a Monte-Carlo. 

In the next step the replicated values $X^R$ and $\tilde\nu^R$ are treated as if they were the outcome of a measurement that is entirely independent of the original measurement. For these values we want to compute the test-statistic.
The constrained likelihood fit for the given value $\mu$ as in equation \ref{eq:fit1} is performed to obtain $\hat\nu_\mu^R$. From the unconstrained fit of equation \ref{eq:fit2} we obtain the values $\mu'^R$ and $\hat\nu_{\mu'^R}^R$. Finally, the test-statistic $q_\mu(X^R)$ can be calculated as defined in equation \ref{eq:llr}.

By repeating this procedure for different replicas, we sample the distribution $f(q_{\mu}|\mu,\hat{\nu}_{\mu})$.
In principle, it is possible to determine the distribution with arbitrary precision. However, computational costs increase with the number of considered replicas.

Additionally, the statistical fluctuation of the test-statistic, if the observation follows the background-only ($\mu=0$) prediction, is considered. Consequently, we determine the distribution $f(q_{\mu}|0,\hat{\nu}_0)$ analogue to $f(q_{\mu}|\mu,\hat{\nu}_{\mu})$. The only difference being that the replicas are generated from the prediction $\lambda(0,\hat{\nu}_0)$.


Another computationally less expensive alternative to obtain the distribution  $f(q_{\mu}|\mu,\hat{\nu}_{\mu})$ is to use an approximate analytical expression as described in \cite{Cowan2011}.  This approach, referred to as the Asimov method, is based on results of \cite{Wilks1938,Wald1943} stating that for a large number of samples  $N$ in the measurement $X$ the likelihood ratio can be approximated by
\begin{equation}
q_{\mu}(X) \sim \left(\frac{\mu - \mu'}{\sigma}\right)^2 + \mathcal{O}\left(\frac{1}{\sqrt{N}}\right).
\end{equation}
The assumption underlying this approximation is that $\mu'$ is Gaussian distributed around its true mean $\mu_0$ with a standard deviation $\sigma$. We obtain the standard deviation $\sigma$ by using
\begin{equation}
	\sigma^2 \approx \frac{(\mu-\mu_0)^2}{q_{\mu}(X^A)},
\end{equation}
where $q_{\mu}(X^A)$ is the test-statistic evaluated for the so-called Asimov set. 
The Asimov set is defined as a set with infinite statistics corresponding to the prediction $\lambda(\mu,\hat{\nu}_{\mu})$. Therefore, observables evaluated using the Asimov set will equal their true values, i.e. $\mu' = \mu_0$. We can determine the Asimov set by either calculating the expected values of the hypothesis exactly or approximate it by performing a sufficiently high statistics Monte-Carlo simulation of the prediction.

The test-statistic can then be shown \cite{Cowan2011} to follow a non-central $\chi^2$-distribution
\begin{equation}
	f(q_{\mu}|\mu,\hat{\nu}_{\mu}) = \frac{1}{2\sqrt{2\pi q_{\mu}}}\left(e^{-\frac{1}{2}\left(\sqrt{q_{\mu}} + \sqrt{\Lambda}\right)^2} + e^{-\frac{1}{2}\left(\sqrt{q_{\mu}} - \sqrt{\Lambda}\right)^2} \right),
\end{equation}
with
\begin{equation}
\Lambda =  \left(\frac{\mu - \mu_0}{\sigma}\right)^2.
\end{equation}
Similarly, it is possible to obtain the distribution $f(q_{\mu}|0,\hat\nu_0)$ using the Asimov method.

\subsection{The $\mathbf{CL_S}$ value and exclusion limits}
Given a measurement $X$ and the corresponding observed value $q_{\mu}(X)$ of the test-statistic as well as the corresponding  distributions $f(q_{\mu}|\mu,\hat{\nu}_{\mu})$ and $f(q_{\mu}|0,\hat{\nu}_0)$, the statistical significance of the observation, i.e. whether it arose by chance, needs to be determined. For that purpose, the $p$-value with respect to the prediction is evaluated:
\begin{eqnarray}
CL_{S+B}(\mu) = \int_{q_{\mu}(X)}^{\infty} dq_{\mu} f(q_{\mu} | \mu,\hat{\nu}_{\mu}) \label{eq:clspb}
\end{eqnarray}
The $p$-value $CL_{S+B}$ is the cumulative probability of observing a measurement $X'$ with $q_{\mu}(X') \geq q_{\mu}(X)$, assuming that the prediction $\lambda(\mu,\hat\nu_\mu)$ correctly describes the measurement outcome. Therefore, large values of $CL_{S+B}$ suggest a high chance that the observation is compatible with $\lambda(\mu,\hat\nu_\mu)$.

The probability to observe a measurement that has a larger $q_{\mu}$ than the observed one, if the background-only prediction describes the observation, is given by the $p$-value $1-CL_B$.
\begin{eqnarray}
1-CL_{B}(\mu) = \int_{q_{\mu}(X)}^{\infty} dq_{\mu} f(q_{\mu} | 0,\hat{\nu}_{0}) \label{eq:clb}
\end{eqnarray}
This probability is a measure for the disagreement of $X$ with the background-only prediction $\lambda(0,\hat\nu_0)$.
For Higgs boson searches this quantity gives an insight into how frequently a measurement suggesting a higher signal contribution than $X$ would be obtained by background fluctuations.

We define the ratio
\begin{equation}
CL_{S}(\mu) \equiv \frac{CL_{S+B}(\mu)}{1-CL_{B}(\mu)}.
\label{eq:cls}
\end{equation}
as measure of how well $\lambda(\mu,\hat\nu_\mu)$ can be statistically distinguished from $\lambda(\mu',\hat\nu_{\mu'})$ based on $X$. In the literature, this is referred to as $CL_S$ confidence level. 
Small values of $CL_S(\mu)$ suggest that $X$ strongly favors $\lambda(\mu',\hat\nu_{\mu'})$ over $\lambda(\mu,\hat\nu_\mu)$. Normalizing to $1-CL_{B}$ ensures sufficiently confident statements, even in the case that fluctuations of the background-only prediction are likely to be similar to the best-fit prediction $\lambda(\mu',\hat\nu_{\mu'})$.

We want to find the minimum $\mu^{\alpha\%}$ of the parameter of interest corresponding to a prediction $\lambda(\mu^{\alpha\%},\hat\nu_{\mu^{\alpha\%}})$ that can be distinguished from $\lambda(\mu',\hat\nu_{\mu'})$ at an $\alpha\%$ $CL_S$ confidence level. To achieve this, we invert the relation $CL_S(\mu) = 1 - \alpha\%$  \cite{Aad2011}, such that
\begin{equation}
\mu^{\alpha\%} = CL_{S}^{-1}(1-\alpha\%).
\end{equation}

Given some measurement $X$, we can define a prediction of the model with $\mu < \mu^{\alpha\%}$ to be indistinguishable from the best-fit prediction at an $\alpha\%$ $CL_S$ confidence level.  Conversely, a prediction corresponding to $\mu > \mu^{\alpha\%}$ is said to be statistically distinguished and excluded at an $\alpha\%$ $CL_S$ confidence level. $\mu^{\alpha\%}$ is referred to as exclusion limit for the parameter of interest $\mu$.


\section{Nuisance parameters for theoretical uncertainties}
\label{sec:nuisance}

The renormalization and factorization scale uncertainty and parton distribution uncertainty have large effects on cross-section predictions for the LHC.
We include these uncertainties in our analysis using nuisance parameters.
We consider nuisance parameters that directly affect the prediction of the signal cross-section:
\begin{equation}
\sigma_{S} = \sigma_{S}^0 \times \nu_{scale} \times \nu_{pdf}
\end{equation}

$\sigma_{S}^0$ is the signal cross-section predicted when choosing the default values for all parameters, $\nu_{scale}$ is the perturbative uncertainty nuisance parameter and $\nu_{pdf}$ is the parton density nuisance parameter.

To use the method as described above and as it is used by ATLAS and CMS \cite{atlashiggscombo,CMS-PAS-HIG-11-032}, it is necessary to choose likelihoods $\pi_{pdf}(\tilde\nu_{pdf}|\nu_{pdf})$ and $\pi_{scale}(\tilde\nu_{scale}|\nu_{scale})$ for the outcomes of auxiliary measurements of the nuisance parameters.

\begin{figure}[H]
	\centering
	\includegraphics[width=0.9\textwidth]{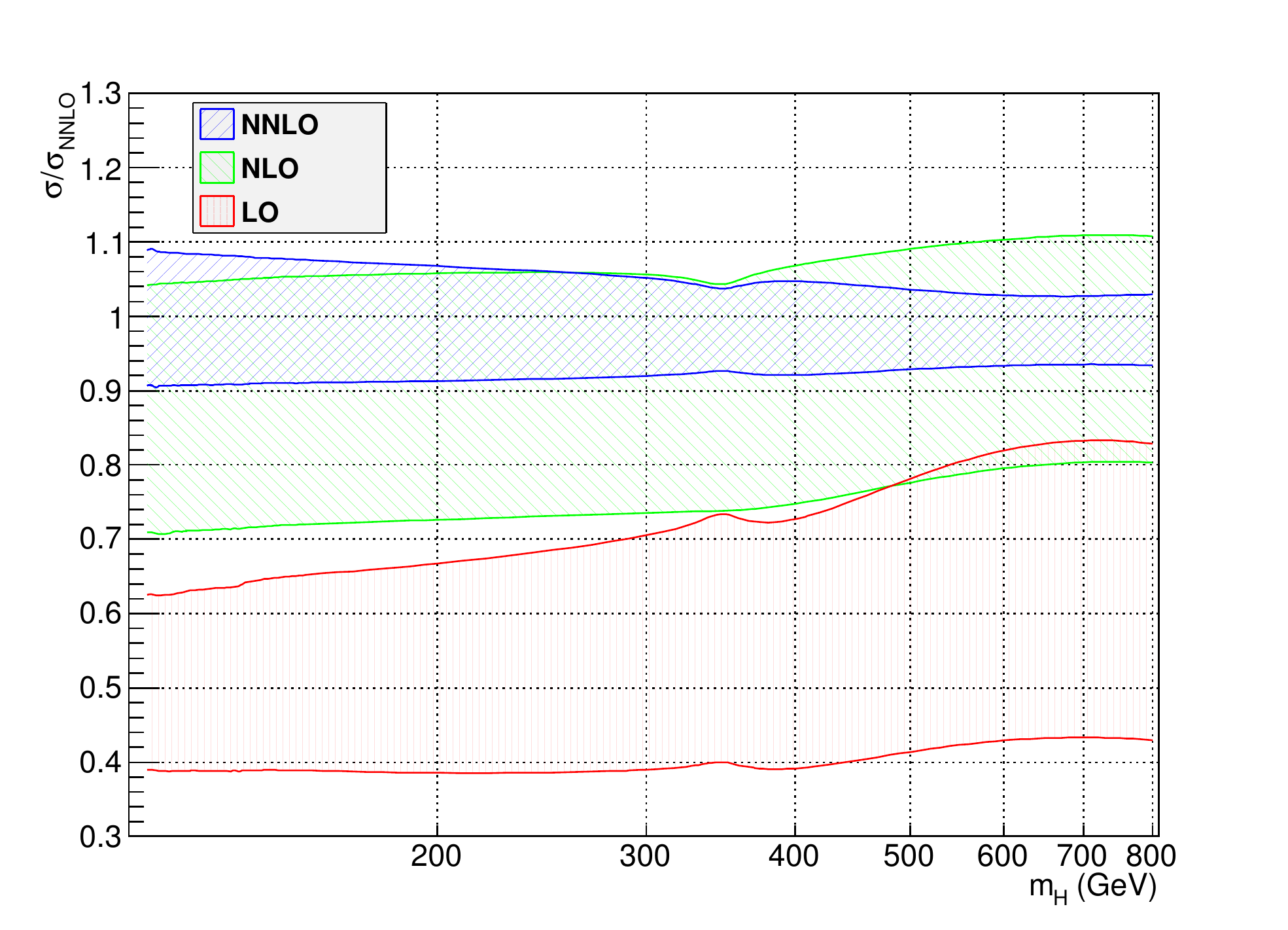}
	\caption{Scale variation of the gluon fusion cross-section at LO, NLO, and NNLO. The LO, NLO and NNLO cross-sections at scale $\mu \in \left[\frac{m_h}{4},m_h\right]$ are normalized to the NNLO cross-section at the central scale $\mu = \frac{m_h}{2}$. Taken from \cite{Anastasiou2012}.}\label{fig:del}
\end{figure}

\begin{description}
\item [Parton density uncertainties] \hfill \\
Partonic predictions need to be convoluted with the parton distributions to obtain physical observables \cite{Demartin2010}. These distributions cannot be computed in the framework of perturbative quantum field theory and need to be measured. The uncertainty on parton distributions arises from various sources including experimental uncertainties of said measurements. The providers of the parton distribution functions give recommendations on how to compute the uncertainty on cross-section predictions for experiments \cite{Ball2010,Lai2010,Martin2009,Watt2012}, usually resulting in an interval $[\sigma_S^-,\sigma_S^+]$ of cross-section predictions.

ATLAS and CMS \cite{atlashiggscombo} model the uncertainty due to the parton densities by using a log-normal distribution
\begin{equation}
\pi_{pdf}(\tilde\nu_{pdf}|\nu_{pdf})=\frac{1}{\sqrt{2\pi} \delta_{pdf}} \frac{1}{\tilde\nu_{pdf}} \exp\left(-\frac{(\ln\tilde\nu_{pdf}-\ln\nu_{pdf})^2}{2\delta_{pdf}^2}\right),
\end{equation}
where $\delta$ is computed from the derived uncertainty on the cross-section $\sigma_S^\pm=\sigma_S^0\times\nu^\pm_{pdf}$
\begin{equation}
\delta_{pdf} = \frac{1}{2} \left(\nu^+_{pdf} - \nu^-_{pdf}\right).
\end{equation}
Using a log-normal distribution rather than a normal distribution ensures that the predictions calculated from varying the nuisance parameter cannot become negative. 

\item [Perturbative uncertainties] \hfill\\
Perturbative uncertainties are the consequence of missing higher order corrections. For Higgs processes, \cite{Anastasiou2012,Anastasiou2002} choose $\mu_F = \mu_R  = Q=\frac{m_h}{2}$ as a most likely value for the renormalization and factorization scale and estimate the uncertainty on the cross-section by variation of $Q$ in the interval $[\frac{m_h}{4},m_h]$.
This variation of the scale allows to estimate the uncertainty on the predicted value of the cross-section. The interval in between the minimal and maximal variation of the cross-section is denoted by $\left[\sigma_S^0\times\nu^-_{scale},\sigma_S^0\times\nu^+_{scale}\right]$. In our opinion, every value within this range is equally plausible, as no independent measurements of the Higgs cross-section have been carried out in the past. Consequently, we do not consider any value within this range to be preferred.
The central value of the interval does in general not correspond to the preferred scale choice $\mu = \frac{m_h}{2}$ and is of no special significance.

This is supported by studies performed by \cite{Anastasiou2012}, which can be seen in figure \ref{fig:del}. The plot shows a comparison of the LO, NLO and NNLO results for gluon-fusion production of a Higgs together with their corresponding theory uncertainties. As one can see, the error-intervals of the different orders are not centered around the same cross-section even though the spread of the intervals decreases with higher orders.
 To treat the values in the interval equally we model $\nu_{scale}$ with a uniform likelihood:
\begin{equation}
\pi_{scale}(\tilde\nu_{scale}|\nu_{scale}) = \frac{1}{\nu^+_{scale}-\nu^-_{scale}}\left(\theta\left(\nu_{scale}-\nu^-_{scale}\right)-\theta\left(\nu_{scale}-\nu^+_{scale}\right)\right)
\label{eq:fp}
\end{equation}
This was also discussed in \cite{Anastasiou2009,Baglio2010,Baglio2011hc}. Here $\theta$ is the Heaviside function.
The right-hand side of equation \ref{eq:fp} is independent of $\tilde{\nu}_{scale}$, as we cannot define a fluctuation of the auxiliary measurement of the perturbative uncertainty and therefore eliminate $\tilde{\nu}_{scale}$ by setting it to unity.

ATLAS and CMS incorporate the perturbative uncertainty differently and describe $\nu_{scale}$ using a log-normal distribution \cite{atlashiggscombo}, analogue to the parton density uncertainty,
\begin{equation}
\pi_{scale}(\tilde\nu_{scale}|\nu_{scale})=\frac{1}{\sqrt{2\pi} \delta_{scale}} \frac{1}{\tilde\nu_{scale}} \exp\left(-\frac{(\ln\tilde\nu_{scale}-\ln\nu_{scale})^2}{2\delta_{scale}^2}\right)
\end{equation}
$\delta_{scale}$ is related to our definition of the relative variation of the cross-section via
\begin{equation}
\delta_{scale} = \frac{1}{2} \left(\nu^+_{scale} - \nu^-_{scale}\right)
\end{equation}
We point out that this distribution introduces a preference for the central values of the interval. Furthermore, it is possible to find a fit value for $\nu_{scale}$ that is outside of the range  $\left[\nu^-_{scale},\nu^+_{scale}\right]$ considered reasonable.
\end{description}
 
The interpretation of experimental results is sensitive to the modeling of nuisance parameters and their specific uncertainties. Therefore, various statistical approaches have been devised, to systematically treat them in limit-setting procedures \cite{Junk1999,Lyons2008,Gross:1099994,ReadCLS}.

\begin{description}

\item[The profiling method]\hfill \\
This method is used by the collaborations ATLAS and CMS  \cite{atlashiggscombo,CMS-PAS-HIG-11-032} and we described it in our review of the $CL_S$ method above.
In the profiling method, the nuisance parameters are determined by performing a constrained maximum-likelihood fit, see equation \ref{eq:fit1}, for a given parameter of interest of the prediction $\lambda$ to the observed data. This retains only the functional dependence of $\lambda$ on $\mu$, while $\nu$ is set to the corresponding optimal value $\hat{\nu}_{\mu}$. Therefore, for a given $\mu$ the agreement of $\lambda$ with the observed data $X$ is maximized. Fitting nuisance parameters in likelihood based analyses is discussed in more detail in \cite{Reid2010,Murphy2000}.
\item[The worst-case method]\hfill \\
The simplest approach is to fix a nuisance parameter to a single value that is considered possible when generating the replicas or the Asimov set and evaluating the test-statistic. This method biases the resulting parameter of interest towards the value favored by that choice. By fixing $\tilde{\nu}$ and $\nu$ to unity for our nuisance parameters, the uncertainty on the nuisance parameter is effectively ignored.

The so-called worst-case method chooses a value for $\nu$ from the allowed range and sets $\pi(\tilde{\nu}|\nu) = 1 \ \forall \ \tilde{\nu}$, such that the result of the analysis is maximally biased towards weaker constraints on the parameter of interest. This corresponds to maximally reducing the predicted signal cross-section. 

\end{description}

We claim that the worst-case method should be used to treat the perturbative uncertainty to obtain physically reasonable limits. To show the impact of the different approaches to treat nuisance parameter we compute exclusion
limits for a toy Higgs search. We show that using a uniform distribution for $\pi_{scale}(\tilde\nu|\nu)$ yields limits similar to those obtained from using the worst-case method.

\subsection*{Setup of the simulated Higgs search}
We conduct a toy Higgs search using the analysis prescriptions outline before. We simulate idealized measurements using a parton-level Monte-Carlo of proton-proton collisions at a center of mass energy of 8TeV, completely neglecting detector effects.
We focus on the $h\rightarrow\gamma\gamma$-channel, investigating a Higgs boson mass range of 110GeV to 140GeV. We choose the invariant mass distribution of the di-photon system, $\frac{\partial N}{\partial m_{\gamma\gamma}}$, as observable for the $CL_S$ method.

To acquire the prediction for the irreducible background, we use the event generator Pythia8 \cite{Sjostrand2008}, collecting $\mathcal{O}(10^7)$ events which we scale to an integrated luminosity corresponding to  $10\text{fb}^{-1}$.
To comply with NNLO results, we rescale the resulting distribution with a  k-factor computed by \cite{Catani2011}.
We simulate the signal prediction using iHixs \cite{Anastasiou2012,Anastasiou2011} at NNLO with gluon fusion as production mode, for every 1GeV bin in the mass range considered.
We combine the background prediction with the signal for a Higgs mass of 125 GeV in order to simulate a measurement $X$.

The perturbative and parton density uncertainties of the signal prediction can be obtained directly from iHixs. We estimate the perturbative uncertainty using the prescription above determining the maximum allowed upward and downward variation $\nu^{\pm}_{scale}$.
iHixs determines the parton distribution uncertainty using the $90\%$ confidence interval of the MSTW08 set, as argued by \cite{Anastasiou2012}.
We then use the variation $\nu^{\pm}_{pdf}$ of the cross-section observed in the process to determine $\delta_{pdf}$ of the log-normal distribution used for $\pi_{pdf}(\tilde\nu_{pdf}|\nu_{pdf})$.

\subsection*{Effect of nuisance parameters on the $CL_S$ method}
For the purpose of this section we consider only the effect of the perturbative uncertainty on the signal cross-section and drop the index $_{scale}$.
We compare the results obtained from treating a flat distribution for the perturbative uncertainty using the profiling method with the results obtained from using the worst-case method.
Additionally we consider the profiling method treating the perturbative uncertainty with a log-normal likelihood $\pi(\tilde\nu|{\nu})$ for the auxiliary measurement in accordance with the prescription used by ATLAS and CMS  \cite{atlashiggscombo}. We set $\tilde{\nu}$ to the default value one. We neglect the condition $\mu\geq\mu'$ and analyze our simulated observation at a chosen Higgs mass point of $m_h=125$GeV, i.e. on the resonance.

The constrained fit (see equation \ref{eq:fit1}) in the calculation of the likelihood using the profiling method tries to improve the agreement of the prediction $\lambda(\mu,\hat\nu_\mu)$ with our simulated data $X$.
The fit chooses $\hat{\nu}_{\mu}$ such that $\lambda$ is increased for $\mu\leq\mu'$ and decreased for $\mu\geq\mu'$.
If $\pi$ is uniform the nuisance parameter can easily take on extremal values, as all values in the considered range are given the same weight. If $\pi$ is log-normal, the values in the tails are suppressed and the nuisance parameter is biased to take values close to one.

\begin{figure}[H]
	\centering
	\subfigure[ ]{\includegraphics[width=0.45\textwidth]{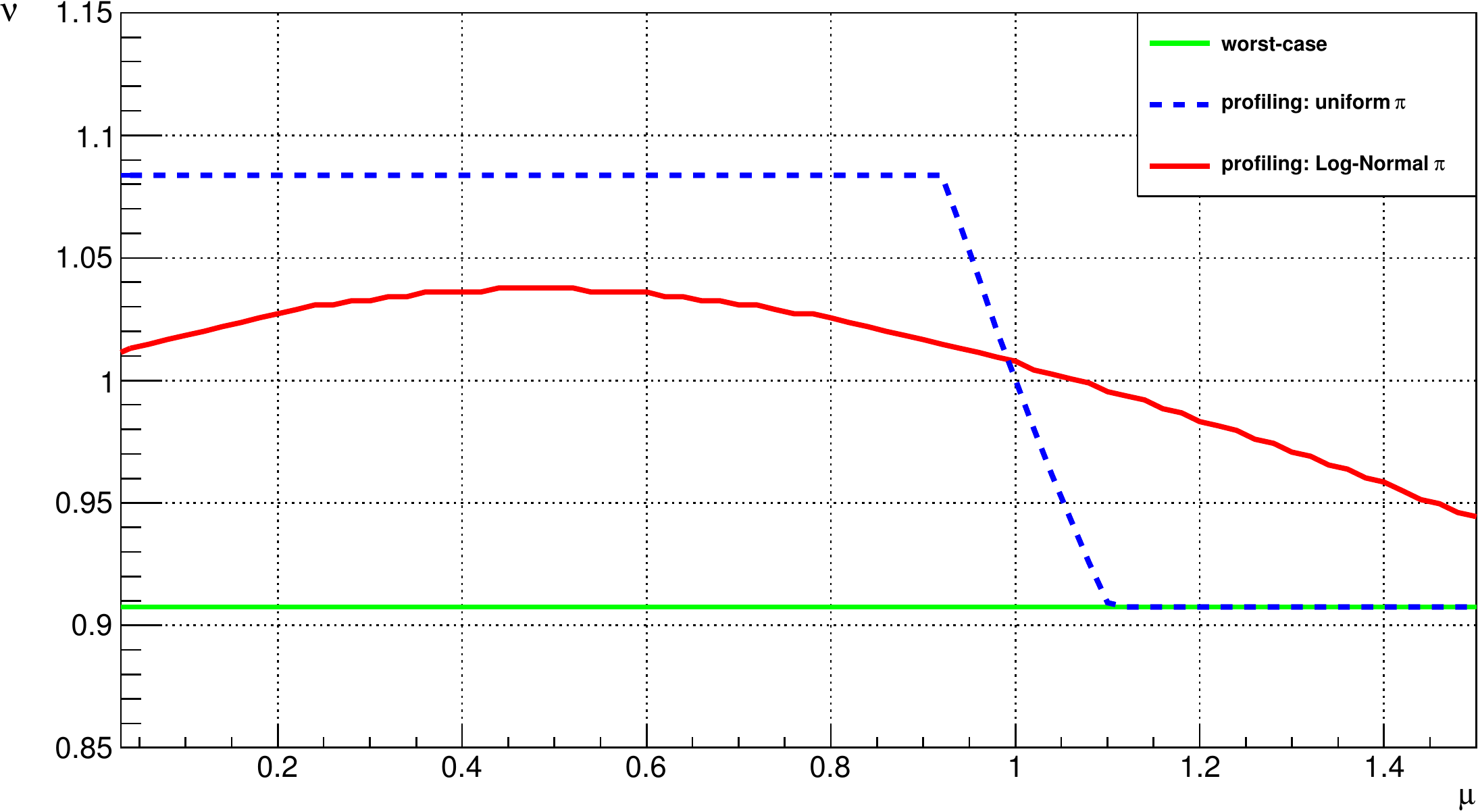}\label{fig:nu}}
	\subfigure[ ]{\includegraphics[width=0.45\textwidth]{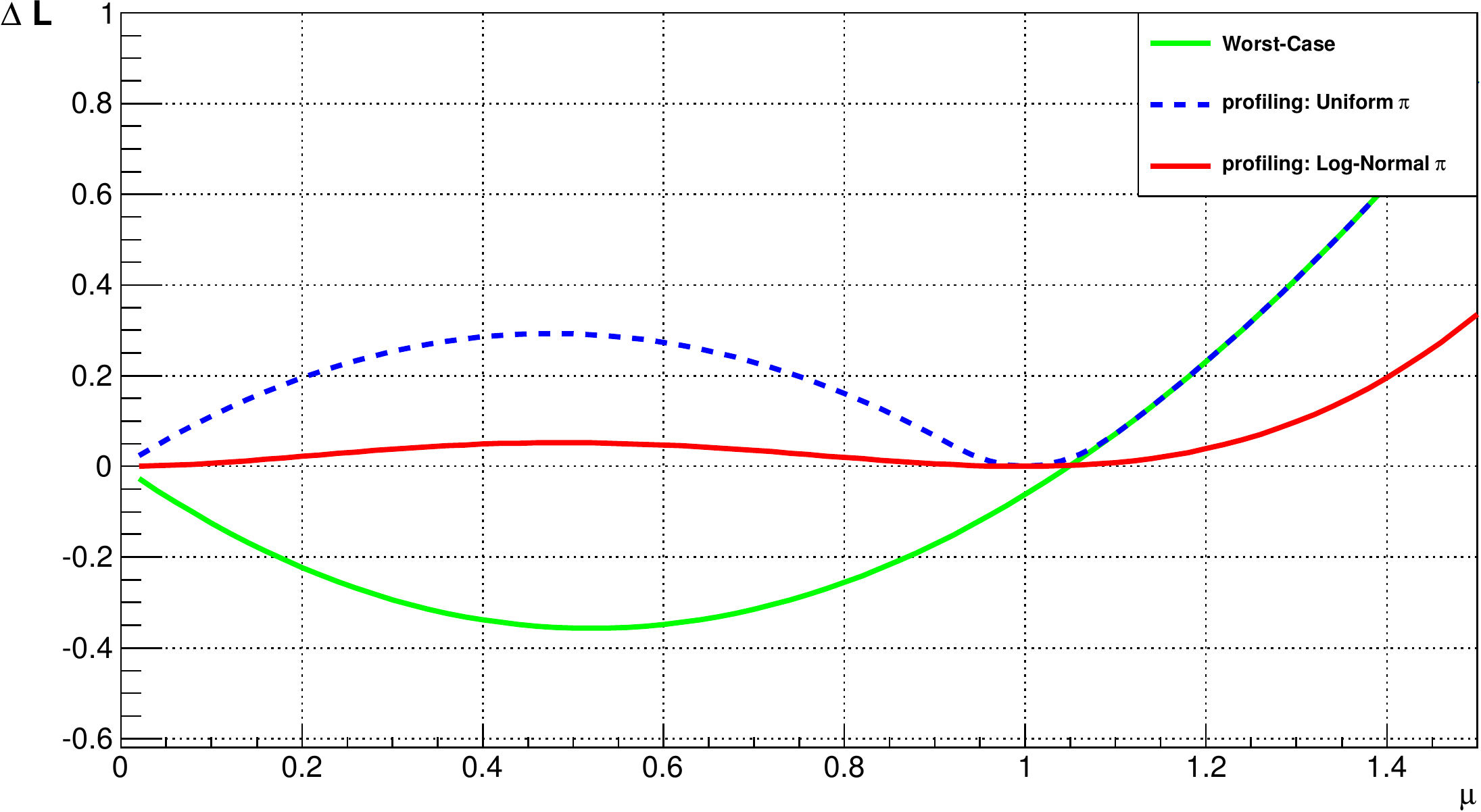}\label{fig:deltaL}}
	\vspace{-0.5cm}
	\caption{a) Nuisance parameter value as function of $\mu$ b) Difference of the likelihood obtained from the specific method to the likelihood obtained from setting the parameters to their default values. The results were obtained with the worst-case (green) and the profiling method using a uniform (blue) or log-normal (red) likelihood $\pi(1|\hat\nu_\mu)$. One can see that the worst-case method and the profiling method with a flat distribution result in equal nuisance parameter values for $\mu > \mu'$. Consequently, we obtain the same likelihood values from both methods in that region.}
\end{figure}

For the worst-case method the nuisance parameter is set to the smallest allowed value. For $\mu$ sufficiently larger than $\mu'$ the profiling method with uniform $\pi$ and the worst-case method result in the same nuisance parameter value.  This is illustrated in Figure \ref{fig:nu}.

The likelihood for a given value $\mu$ is increased if the nuisance parameter is fitted to the observation compared to the likelihood if $\nu$ is fixed to unity. As the uniform $\pi$ does not give different weights to different parameter values in the allowed range the increase is larger than in the log-normal case. The worst-case and the profiling method with uniform $\pi$ yield the same nuisance parameters for sufficiently large $\mu$, the likelihoods are therefore equal in that region. This can be seen in Figure \ref{fig:deltaL}.

\begin{figure}[h]
	\centering
	\includegraphics[width=0.9\textwidth]{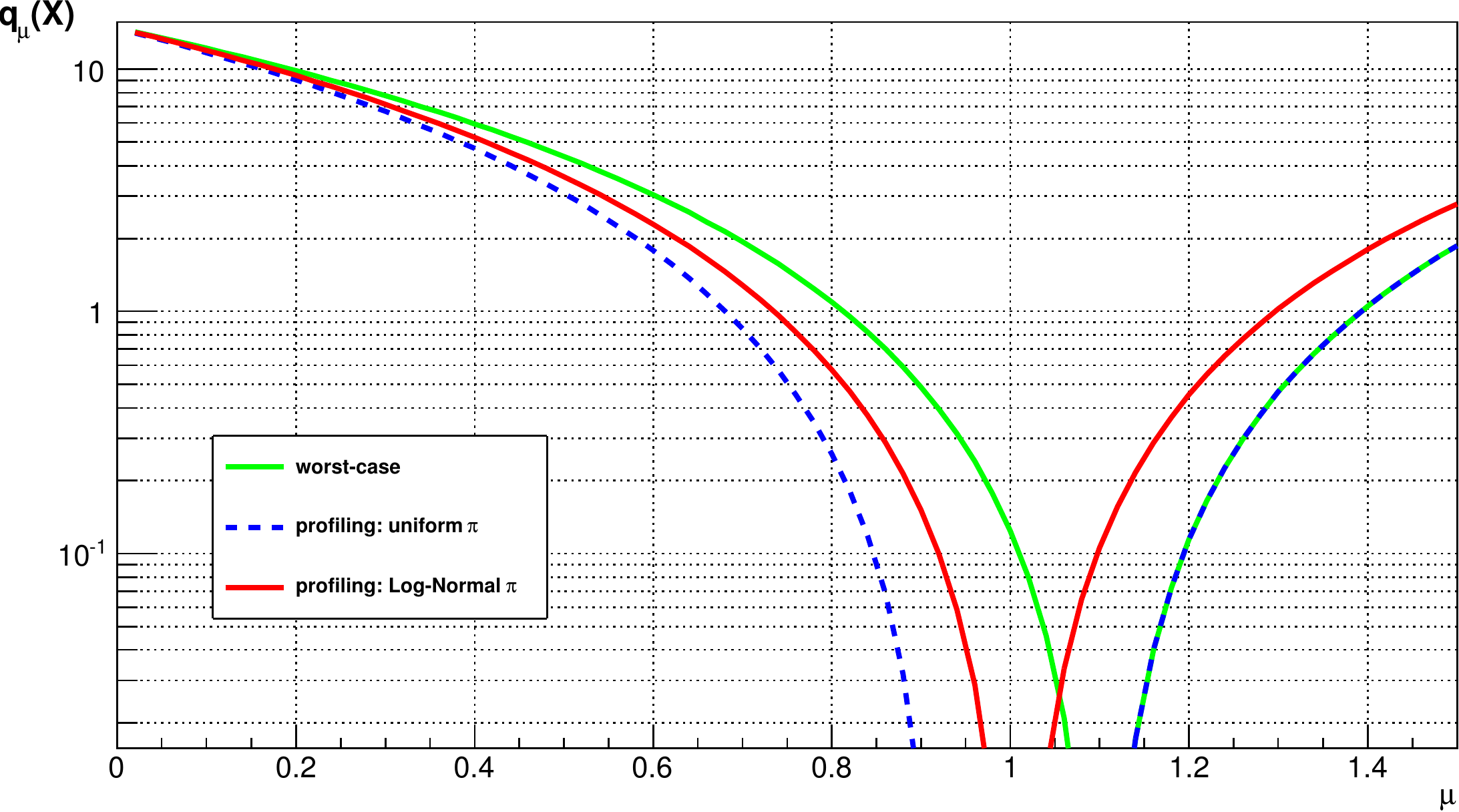}
	\caption{The test-statistic obtained with the worst-case (green) and the profiling method using a uniform (blue) or log-normal (red) likelihood $\pi(1|\hat\nu_\mu)$ as a function of $\mu$. As one can see, the worst-case method and the profiling method with a flat $\pi$ result in equal values for the test-statistic in the region $\mu>\mu'$.  Consequently, the exclusion limits obtained from both methods are the same.}\label{fig:llr}
\end{figure}

Next, we compute the test-statistic $q_\mu(X)$ using the profiling and worst-case methods, see equation \ref{eq:llr}. Profiling nuisance parameters reduces the difference between the best-fit prediction and the prediction $\lambda(\mu|\hat\nu_\mu)$ corresponding to the considered $\mu$. Consequently, $q_\mu(X)$ is reduced for every $\mu$. The effect is again largest for profiling with a uniform $\pi$. In the vicinity of $\mu\approx\mu'$ the nuisance parameter can compensate for small deviations of the predicted cross-section from the best-fit value and $q_\mu(X)$ is compatible with zero. For the worst-case method $q_\mu(X)$ is also minimal for $\mu=\mu'$ but the best-fit value $\mu'$ is larger than in the profiling method due to the reduced signal cross-section in the prediction. Again, for $\mu$ sufficiently larger than $\mu'$ the worst-case method and the profiling method with uniform $\pi$ give the same results. This is depicted in Figure \ref{fig:llr}.

\begin{figure}[h]
	\centering
	\includegraphics[width=0.9\textwidth]{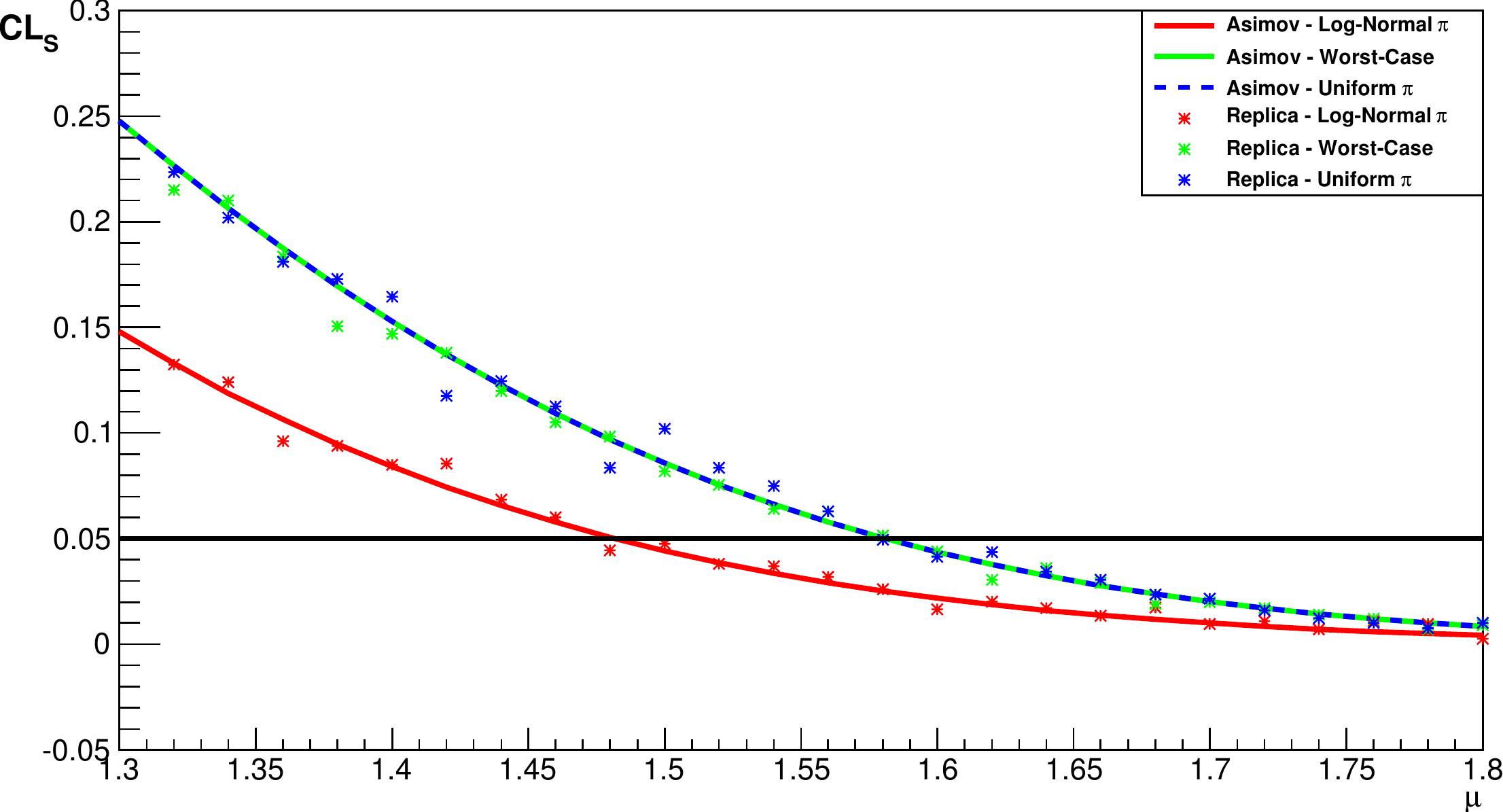}
	\caption{$CL_S$ as a function of $\mu$ obtained with the Asimov (lines) and replica method (dots). The predicted nuisance parameter value was obtained using the profiling method with a uniform $\pi$ (blue), a log-normal $\pi$ (red) and using the worst-case method (green). The exclusion limit $\mu^{95\%}$ is the value of $\mu$ corresponding to $CL_S=0.05$ (black line). One can see that both the profiling of a uniform $\pi$ as well as the worst-case method lead to a higher, i.e. more conservative, exclusion limit in comparison to the profiling of the log-nornal $\pi$.}\label{fig:cls}
\end{figure}

Re-imposing the constraint $\mu\geq\mu'$ we can compute the distributions of $q_{\mu}(X)$ that are necessary to derive the $CL_S$ value, see section \ref{sec:fpdf}. We compute the $CL_S$ values using the worst-case method and compare them to the $CL_S$ values that we obtain when profiling a uniform distribution as well as a log-normal distribution. For each calculation, the $f$ distributions, which are needed to determine $CL_S$, are evaluated using the replica method as well as the Asimov method.
Figure \ref{fig:cls} shows the different $CL_S$ values as functions of $\mu$. As one can easily see, the values obtained using the Asimov and the replica method agree well for both the profiling and the worst-case method, thanks to the large statistics of our simulation. Furthermore one can see that the exclusion limit $\mu^{95\%}$, i.e. the point where $CL_S$ is smaller than $0.05$, is increased when using a uniform $\pi$ in comparison to using a log-normal $\pi$. Therefore, modeling the perturbative uncertainty with the worst-case method leads to conservative exclusion limits.

It should be noted that even though the tails of the log-normal distribution are strongly suppressed it is in principle possible to find outliers for the nuisance parameter value in large signal-to-background ratio scenarios. As pointed out before these can be outside of the range of plausible values for the perturbative uncertainties. We do not consider this behavior to be appropriate for this specific uncertainty. Therefore, we recommend to use the worst-case method as defined above.

\section{Discussion of search results}
\label{sec:combo}
\subsection*{Exclusion limits}
The exclusion limits obtained using the $CL_S$ method described in section \ref{sec:cls} are usually presented in exclusion plots.
These show the observed limits obtained from an actual experiment at the same time as the expected limits.

We compute the observed limits $\mu^{95\%}_{obs}$ by applying the $CL_S$ method to a simulated measurement $X$. In order to quantify the sensitivity of the analysis, we determine the exclusion limits obtained from the background only prediction. We produce 5000 replicas (see section \ref{sec:fpdf}) of the background-only prediction $\lambda(0,\hat{\nu}_0)$. The exclusion limit $\mu_R^{95\%}$ for each of the replicas is subsequently computed in the same way as for the simulated measurement using the $CL_S$ method. This set of $\mu^{95\%}_R$ values samples the underlying probability density $m(\mu^{95\%}_R)$, which contains information about the statistical fluctuations of the expected exclusion limits. We calculate the median expected limit $\mu_{med}^{95\%}$ using
\begin{equation}
	\frac{1}{2} = \int_{-\infty}^{\mu_{med}^{95\%}} d \mu^{95\%}_R m(\mu^{95\%}_R).
\end{equation}
To quantify the range of possible statistical deviations from the median we additionally compute the corresponding $\pm1\sigma$ and $\pm2\sigma$ intervals $[\mu_{-N\sigma},\mu_{+N\sigma}]$ using
\begin{equation}
	\frac{1}{2}(1\pm p_N) = \int_{-\infty}^{\mu_{\pm N\sigma}} d \mu^{95\%}_R m(\mu^{95\%}_R),
\end{equation}

\begin{figure}[H]
	\centering
	\subfigure[]{\includegraphics[width=0.45\textwidth]{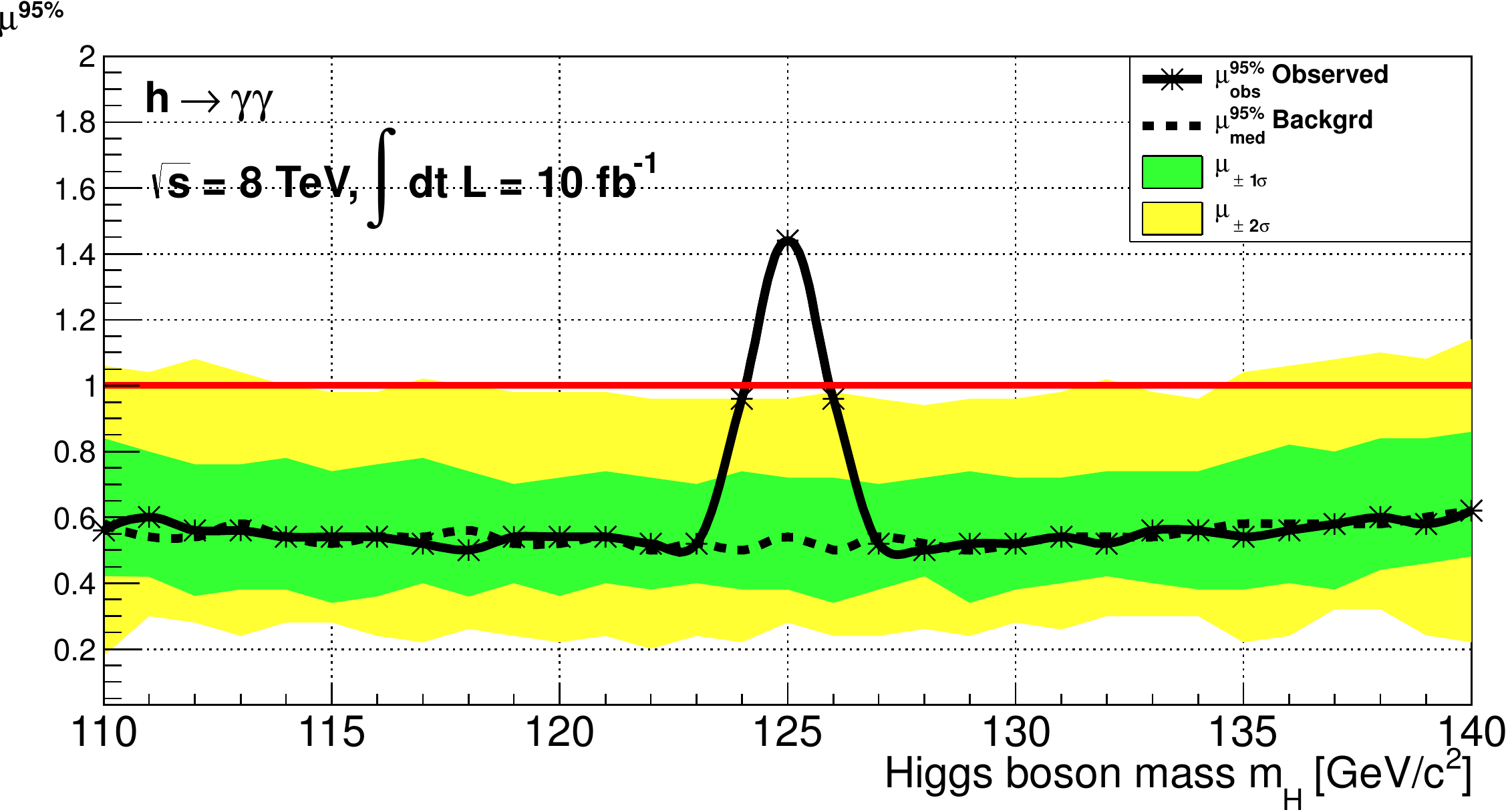}\label{fig:excl}}
	\subfigure[]{\includegraphics[width=0.45\textwidth]{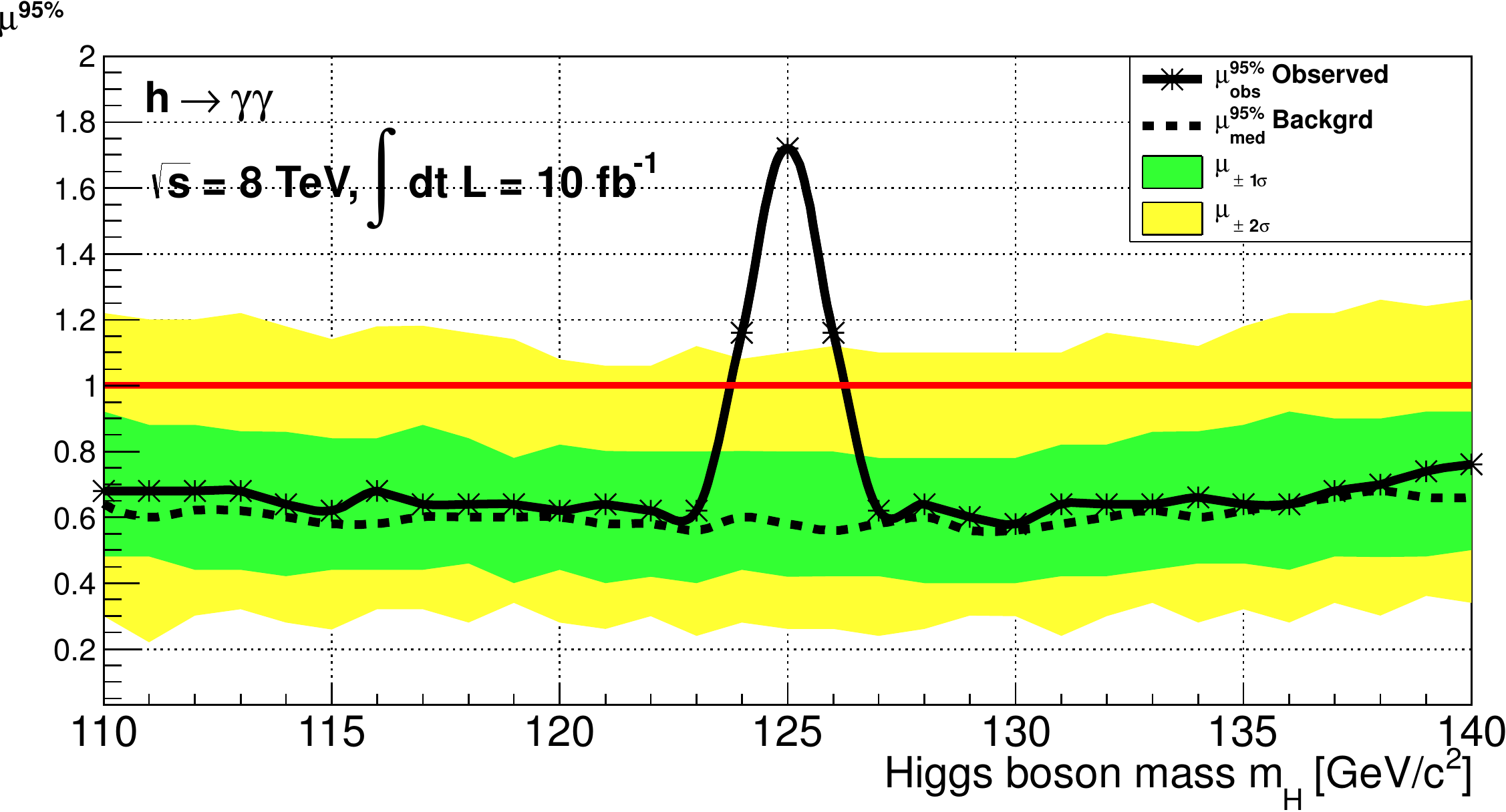}\label{fig:exclFF}}
	\caption{Exclusion limits for standard model Higgs with a mass of 125GeV. The solid black line depicts the observed limits. The dashed line represents the median expected limits, while the green (yellow) band shows the interval expected to contain $68\%$ ($95\%$) of the statistical fluctuations of the expected limits. The nuisance parameters are set to unity in the left plot. A constrained fit is performed for the parton density uncertainty and the worst case method was used for the perturbative uncertainty in the right plot. }
\end{figure}

Figures \ref{fig:excl} and \ref{fig:exclFF} show the $95\%$ $CL_S$ confidence limit on the signal strength multiplier $\mu$ as a function of the hypothesized Higgs boson mass. 
As our goal is rather to demonstrate the expected limits of a standard model Higgs boson search than to mimic an actual measurement including statistical fluctuations, we use the Asimov set as "observation". This suppresses the statistical fluctuations of the black line.
The hypothetical $95\%$ $CL_S$ exclusion limit on the standard model Higgs prediction, $\mu =1$ is marked with the red horizontal line.
The dashed line represents $\mu_{med}^{95\%}$. The green and yellow bands mark the corresponding $\pm1\sigma$ and $\pm2\sigma$ intervals. These allow to assess the exclusion potential of the experiment. If the dashed curve is below the red line, the experiment is sensitive to the standard model Higgs boson.
If the dashed line is above the red line, the statistics of the experiment is too low to be able to exclude the standard model Higgs boson at $95\%$ $CL_S$ confidence level, even if it is not present in the observed data.

We exclude a Higgs boson with a cross section larger than $\mu^{95\%}$ depicted by the black solid line at $95\%$ $CL_S$ confidence level based on the measurement. If the black line is below one, the existence of a standard model Higgs boson is excluded at a $CL_S$ confidence level of at least $95\%$.
As long as the observed exclusion limit is within the green or yellow bands, it agrees with a $1\sigma$ or $2\sigma$ fluctuation of the expected limits. Large excursions of the black line below the $2 \sigma$ band can indicate problems with modeling of the background.
Conversely, a departure of the observed line above the $2\sigma$ band and the red line, indicates that it is not possible to exclude the standard model Higgs boson in this mass range at a $95\%$ $CL_S$ confidence level.
However, such an exclusion limit should by no means be used to quantify a possible excess over the standard model. In particular, an exclusion limit above $\mu=1$ does not give conclusive evidence that the cross-sections of the physical processes are enhanced above the standard model.

Figure \ref{fig:excl} shows the exclusion limits obtained when no systematic uncertainties are taken into account. In contrast to that, Figure \ref{fig:exclFF} shows the limits that are found when the systematic uncertainties are considered by fitting of the parton density nuisance parameter and using the worst-case method for the perturbative uncertainty. By comparison, it becomes apparent that the sensitivity of the experiment is decreased due to the uncertainties. The bands of the expected exclusion limits are wider in \ref{fig:exclFF} compared to the bands in \ref{fig:excl}, allowing for larger statistical excursions of the limits. Furthermore, including nuisance parameters results in weaker exclusion limits.

\subsection*{The significance of an excess}
\begin{figure}[H]
	\centering
	\includegraphics[width=1.00\textwidth]{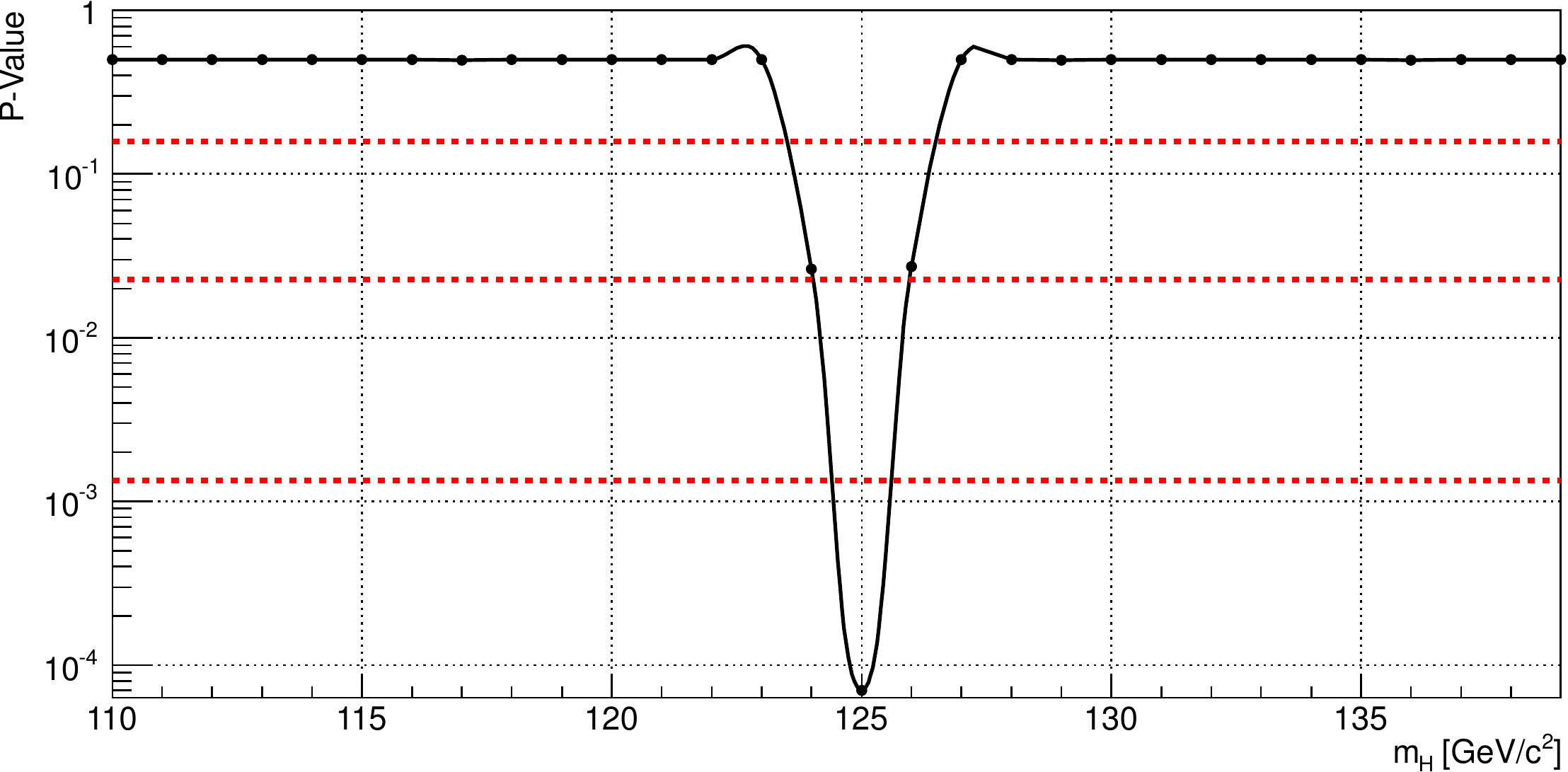}
	\caption{$p$-value plot, visualizing the probability that the background rate fluctuates up higher than the observation as a function of the tested Higgs mass. The red dashed lines correspond to $1\sigma$, $2\sigma$ and $3\sigma$ confidence levels respectively (top to bottom).}
	\label{fig:pval}
\end{figure}

To quantify the significance of an excess, the $p$-value corresponding to the probability that the background could fluctuate up to values larger than the observed value needs to be determined. For that purpose, we use 
\begin{equation}
q_{0}(X) = -2 \log \frac{\mathcal{L}(X|0,\hat{\nu}_0)}{\mathcal{L}(X|\mu',\hat{\nu}_{\mu'})}
\end{equation}
as test-statistic. We determine the observed value for $q_{0}$ in the usual way and estimate the probability density $f(q_0 | 0,\hat{\nu}_0)$ of $q_0$ using the replica method.
We can then visualize the $p$-value 
\begin{equation}
p_0(X) = \int_{q_0(X)}^{\infty} d q_0 f(q_0 |0,\hat{\nu}_0)
\end{equation}
in a separate $p$-value plot, such as figure \ref{fig:pval}. In this plot, smaller $p$-values, i.e. lower probabilities that the observation could be caused by fluctuations of the background, correspond to larger excesses. An excess of at least $5\sigma$ is commonly required to claim a discovery. 

\subsection*{Agreement with the standard model}
\begin{figure}[H]
	\centering
	\includegraphics[width=1.00\textwidth]{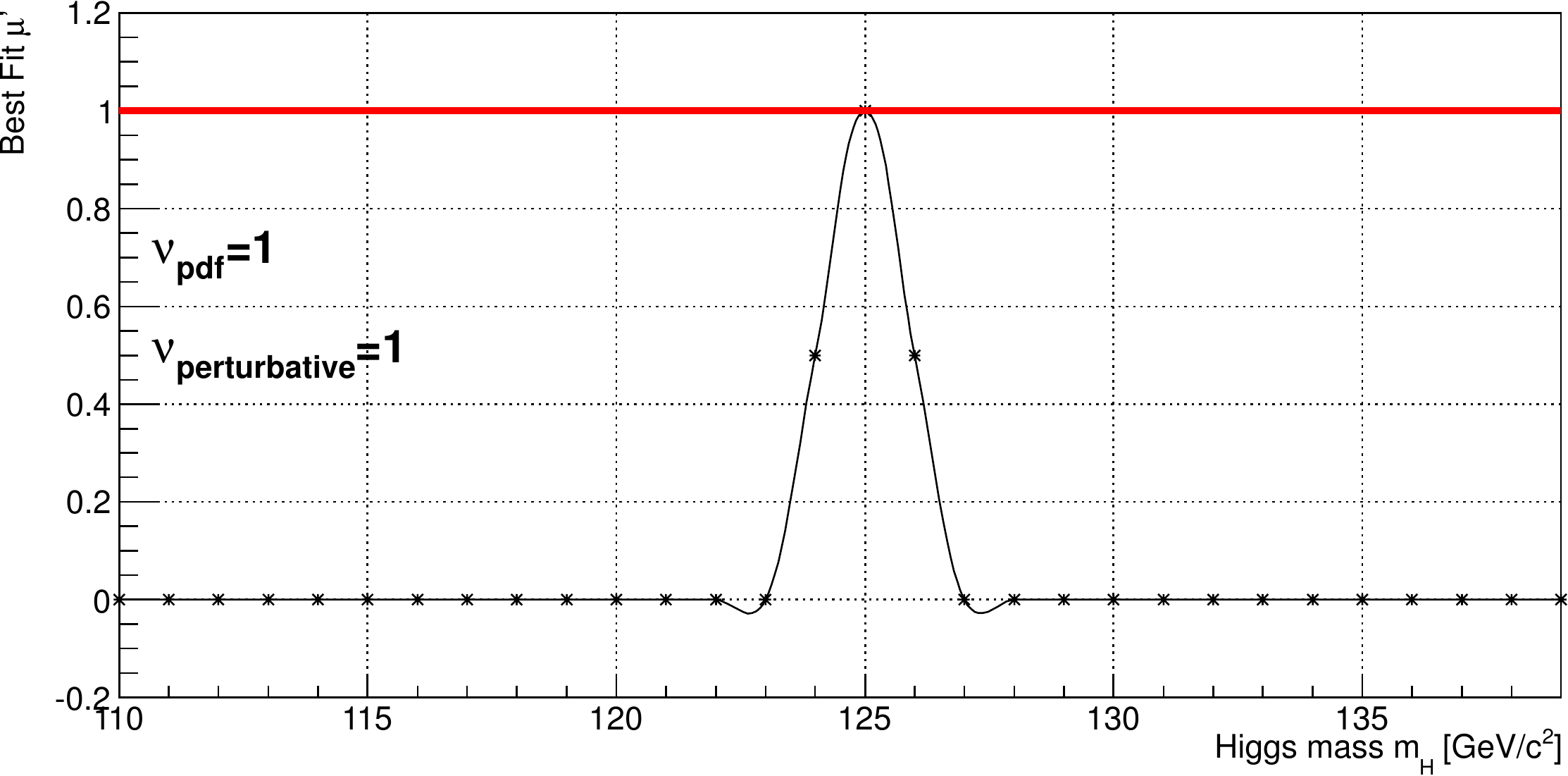}
	\caption{Standard Model agreement plot showing the best-fit value of the signal strength modifier. The excess has a best-fit value of $\mu'=1$, indicating a standard model Higgs with unmodified cross-sections.}
	\label{fig:smagree}
\end{figure}

In order to determine whether an excess of observed events above the background-only prediction indicates enhanced cross-sections, we consider the best fit of the parameter of interest to the observed data. In this analysis we are not so much interested in deriving conservative limits based on the observation but rather want to determine the size of the signal cross-section. To obtain the best fit, we therefore set our nuisance parameters to unity, rather than their worst-case or fitted values, and compute $\mu'$ such that 
\begin{equation}
\mathcal{L}(X|\mu',1)\geq\mathcal{L}(X|\mu,1)\hspace{1cm}\forall\mu.
\end{equation}
The results of this computation are presented in the so-called standard model agreement plot, figure \ref{fig:smagree}. For a standard model Higgs in the measurement, the best fit $\mu'$ should be compatible with one at the corresponding Higgs mass. If it significantly differs from one, this may indicate modified cross-sections.

The treatment and assigned values of the nuisance parameters should be stated along with the plot in order to be able to draw conclusions on theoretical models.

Comparing figures \ref{fig:excl}, \ref{fig:exclFF} and \ref{fig:smagree}, one can see that although the exclusion plots show an excess of $\mu^{95\%} \approx 1.7$ at $m_h=125$GeV, corresponding to a significance of more than $3\sigma$, this does not indicate modified cross-sections, as the corresponding standard model agreement plot shows that the respective best-fit value is compatible with unity.
\section{Conclusion}
We investigated the $CL_S$ technique, which is the method used to set exclusion limits on the contributions of a standard model like Higgs boson to observables. The ATLAS and CMS collaborations utilize the profile LLR to take into account nuisance parameters. To compute limits, both the Asimov and the replica method, are used.

We demonstrated that theoretical uncertainties have a significant impact on the exclusion limits derived with the $CL_S$ method. The parton density function uncertainty was modeled in our analysis as a log-normal distribution and treated using the profiling method. 

ATLAS and CMS currently model the perturbative uncertainty using a log-normal distribution.
Profiling this specific uncertainty and thereby treating it as if it was the consequence of a insufficiently precise measurement of the renormalisation and factorisation scale is not adequate.
Furthermore, our analysis shows that the limits derived using this technique are not conservative as the impact of the perturbative uncertainty is effectively underestimated.
In our opinion, it is necessary to choose the most conservative signal cross-section plausible according to theoretical considerations.
We therefore recommend to incorporate the perturbative uncertainty using the worst-case method.
High-precision analyses will require calculations at higher order in perturbation theory that reduce the perturbative uncertainty.

An investigation of more physical alternatives to the unphysical signal-strength modifier is a topic of future studies.

\acknowledgments
We thank Babis Anastasiou for his essential support, Ben Kilminster for clarification of the current procedures used by ATLAS and CMS, and Alexander Aeberli for useful discussions.
This research is supported by the ERC Starting Grant for the project "IterQCD".

\bibliographystyle{JHEP}  
\bibliography{paper}  

\end{document}